\documentclass[]{raa}            

\usepackage{graphicx,times}             
\usepackage{subfigure}
\usepackage{natbib}
\usepackage{textcomp}
\usepackage{mathtext}
\usepackage{amssymb,amsmath}
\usepackage{longtable,booktabs}

\newcommand{\fermi}{\textit{Fermi}}
\newcommand{\gr}{$\gamma$-ray}

\begin{document}

   \title{Identification of Candidate Millisecond Pulsars from \textit{Fermi} LAT Observations 
}

   \volnopage{Vol.0 (2015) No.0, 000--000}      
   \setcounter{page}{1}          

   \author{Xuejie Dai 
      \inst{1,2}
   \and Zhongxiang Wang
      \inst{1}
\and V. Jithesh
	\inst{1}
   \and Yi Xing
      \inst{1}
   }

   \institute{Shanghai Astronomical Observatory, Chinese Academy of Sciences,
             Shanghai 200030, China; {\it wangzx@shao.ac.cn}\\
        \and
             Graduate University of the Chinese Academy of Sciences, No. 19A, Yuquan Road, Beijing 100049, China\\
   }


\abstract{We report our detailed data analysis for 39 $\gamma$-ray sources
selected from the 992 unassociated sources in the \textit{Fermi}
Large Area Telescope (LAT) third source catalog. 
The selection criteria, which were
set for finding candidate millisecond pulsars (MSPs), are non-variables
with curved spectra and $>$5$^{\circ}$ Galactic latitudes.
From our analysis, 24 sources were found to be point-like sources not
contaminated by background or nearby unknown sources. Three of them,
J1544.6$-$1125, J1625.1$-$0021, and J1653.6$-$0158, have been previously
studied, indicating that they are likely MSPs. The spectra 
of J0318.1+0252 and J2053.9+2922 do not have properties similar to that 
of known $\gamma$-ray MSPs, and we thus suggest that they are not MSPs.
Analysis of archival X-ray data for most of the 24 sources were also conducted.
Four sources were found with X-ray objects in their error circles, and 16
with no detection. The ratios between the $\gamma$-ray fluxes and X-ray fluxes 
or flux upper limits are generally lower than those of the known $\gamma$-ray
MSPs, suggesting that if the $\gamma$-ray sources are MSPs,
none of the X-ray objects are the counterparts. Deep X-ray or
radio observations of
these sources are needed in order to identify their MSP nature.
}

   \authorrunning{X. Dai, Z. Wang, V. Jithesh, \& Y. Xing}            
   \titlerunning{\textit{Fermi} Candidate Millisecond Pulsars}  

   \maketitle

%
%
\section{Introduction}           
\label{sect:intro}

The \textit{Fermi Gamma-Ray Space Telescope (Fermi)}, with its great 
capabilities, has
revolutionized our view of the high-energy, \gr\ sky. Thus far, the detection
of 3033 sources has been reported in the \fermi\ Large Area Telescope
(LAT) third source catalog \citep{3fgl15}, which used 4 yr of science data 
(year 2008--2012) from \fermi\ LAT all-sky monitoring observations. Among 
the sources, most of
them are Active Galactic Nuclei (AGN; \citealt{3fagn15}) and in 
our Galaxy, the prominent 
class is pulsars. According to the second \fermi\ LAT catalog of \gr\ pulsars
and public list of LAT-detected \gr\ pulsars\footnote{https://confluence.slac.stanford.edu/display/GLAMCOG/Public+List+of+LAT-Detected+Gamma-Ray+Pulsars},
161 pulsars have been detected with \gr\ pulsations
and more than 20 new millisecond pulsars (MSPs) have been discovered 
due to \fermi\ LAT detection of them. These results have not only established 
pulsars as the main \gr\ sources in the Galaxy, which has long been suspected
from the surveys of the sky with previous Gamma-Ray telescopes, for example the 
\textit{Compton Gamma-Ray Observatory} \citep{tho08}, but also helped 
significantly improve our studies of the pulsar population, in particular 
the MSPs.

MSPs are $\sim$10$^9$ yr old, fast spinning neutron stars, which have
evolved from low-mass X-ray binaries by accreting from companions
and thus gaining sufficient angular momentum \citep{alp+82,rs82}.
Due to their relatively high efficiency $\eta$ of converting spin-down energy
$\dot{E}$ to \gr\ emission (because $\eta \propto 1/\dot{E}$; e.g., 
\citealt{2fpsr13}) and isotropic distribution in the sky, \fermi\ all-sky
monitoring is a powerful tool for finding candidate new MSPs, although note
that it is extremely difficult to identify them from blind searches for 
pulsation signals in the \fermi\ LAT data (e.g., \citealt{ple+12}).
One important result due to \fermi\ is the discovery of
a significant number of eclipsing MSP binaries, namely black 
widows \citep{fst88} and redbacks \citep{rob13}.  As pointed out 
by \citet{rob13}, the number of such systems is increased by 6-times 
to $\sim$20.  Moreover three redbacks, PSR~J1023+0038 \citep{arc+09},
J1824$-$2452I (in the globular cluster M28; \citealt{pap+13}), and
XSS~J12270$-$4859 \citep{bas+14}, are also known as transitional pulsar 
binaries, which can switch between the states of having an accretion disk 
and being disk-free.  How to explain the presence of these systems and their
formation processes is an interesting question \citep{che+13,ben+14}.

Since approximately one third of \fermi\ LAT sources have not been identified 
or found with associations with any known objects \citep{3fgl15} and 
pulsars are the prominent \gr\ sources, it is conceivable that
a significant number of pulsars are among these un-associated sources.
We have carried out a systematic study of them, aiming to identify
MSPs among them.  In this paper, we report target selection from 
the \textit{Fermi} LAT third source catalog for candidate MSPs (101 sources
were found; see Section~\ref{subsec:ts}) and
results from data analysis for 39 of the selected targets. 
In Section~\ref{subsec:ts}, 
the detailed selection is provided, which is based on the properties
of pulsars learned from \fermi\ studies. We present our analysis of
the LAT data and archival X-ray data for the targets
in Section~\ref{sec:ar}. The results and discussion 
are given in Section~\ref{sec:rd}.
\begin{figure}
   \centering
   \includegraphics[width=0.6\textwidth, angle=0]{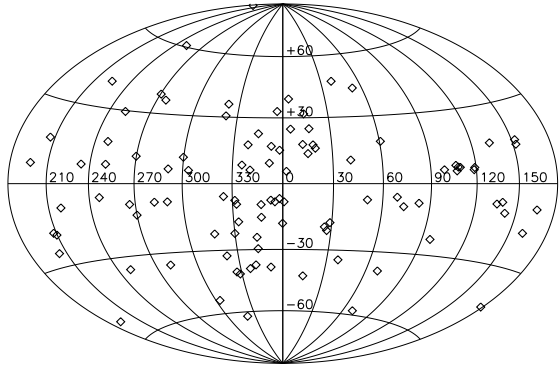}
   \caption{Galactic positions of the selected 101 sources from the LAT third 
source catalog. Nearly 40\% of them are located within Galactic longitudes of
$\pm$30 degrees.}
   \label{fig:candi}
   \end{figure}

\subsection{Candidate target selection}
\label{subsec:ts}

From \fermi\ LAT observations, it has been learned that emission from pulsars
is stable. This feature greatly helps the selection of them from 
the dominant AGN sources. The latter are strong variables at multi-wavelengths 
including \gr\ (e.g., \citealt{wil+14}). 
In addition, the \fermi\ \gr\ spectra of pulsars generally have a form of
an exponentially cutoff 
power law with the cutoff energies at several GeV \citep{2fpsr13}, i.e.,
some degree of curvature in their spectra is one feature of their emission.
For comparison, AGN generally have `straight' power law spectra 
(e.g., \citealt{3fagn15}).

We thus selected candidate MSP targets from the high Galactic sources in 
the LAT third source catalog, since MSPs generally have an isotropic 
distribution \citep{2fpsr13}.
A Galactic latitude of $> 5^{\circ}$ was used, which helped avoid
the crowded Galactic plane.
Then requiring that the sources have variability indices less than 72.44 
(99\% confidence for a source not being a variable)
and the curvature significances greater than 3$\sigma$ \citep{3fgl15},
101 sources were selected from the LAT catalog. Their positions 
in the Galactic coordinates
are shown in Figure~\ref{fig:candi}. As can be seen, 
nearly 40\% of them
are located within Galactic longitudes of $\pm$30 degrees, implying
a high concentration towards the Galactic center direction. Such a
distribution suggests that they are likely associated with the Milky Way.
\begin{figure}
   \centering
   \includegraphics[width=0.4\textwidth, angle=0]{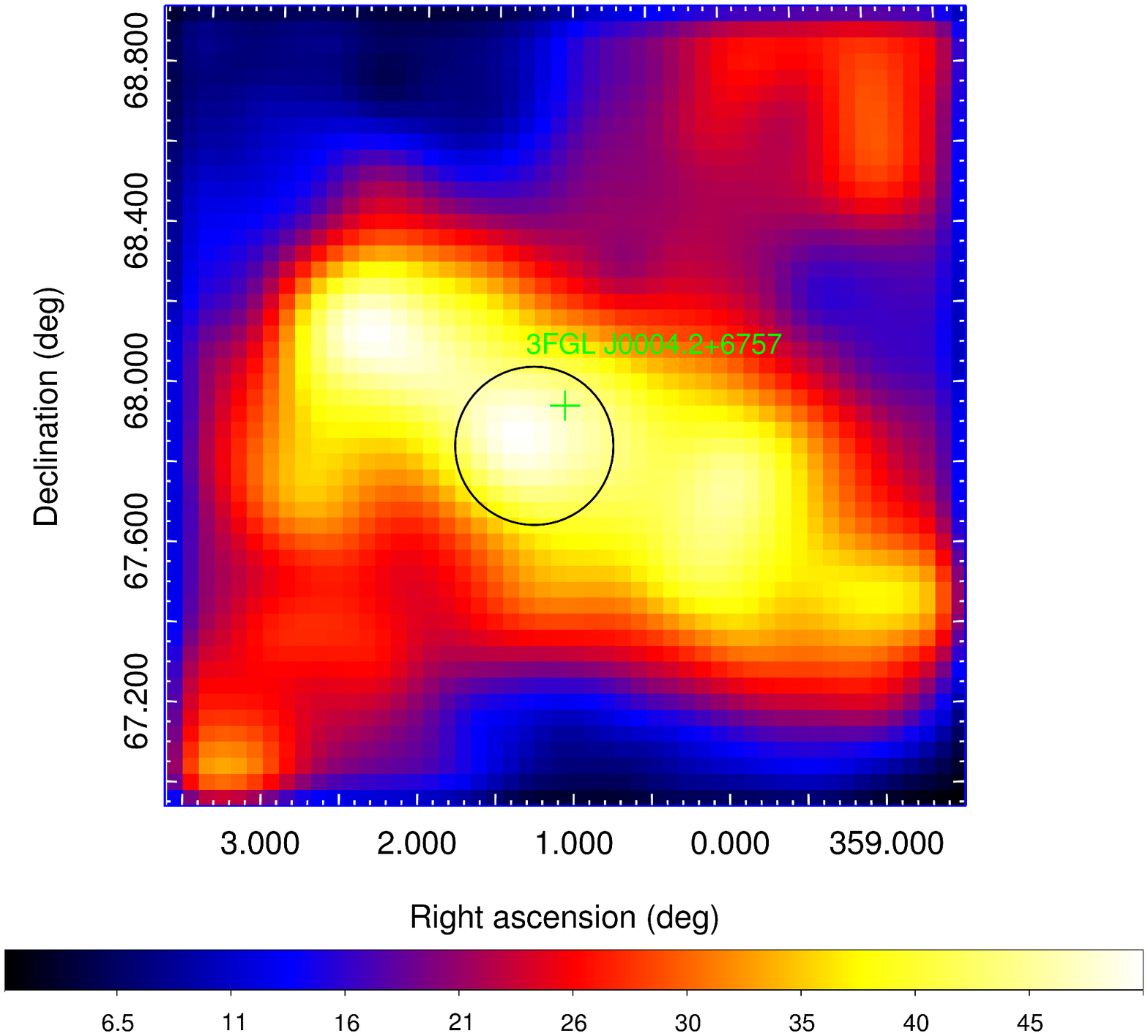}
   \includegraphics[width=0.4\textwidth, angle=0]{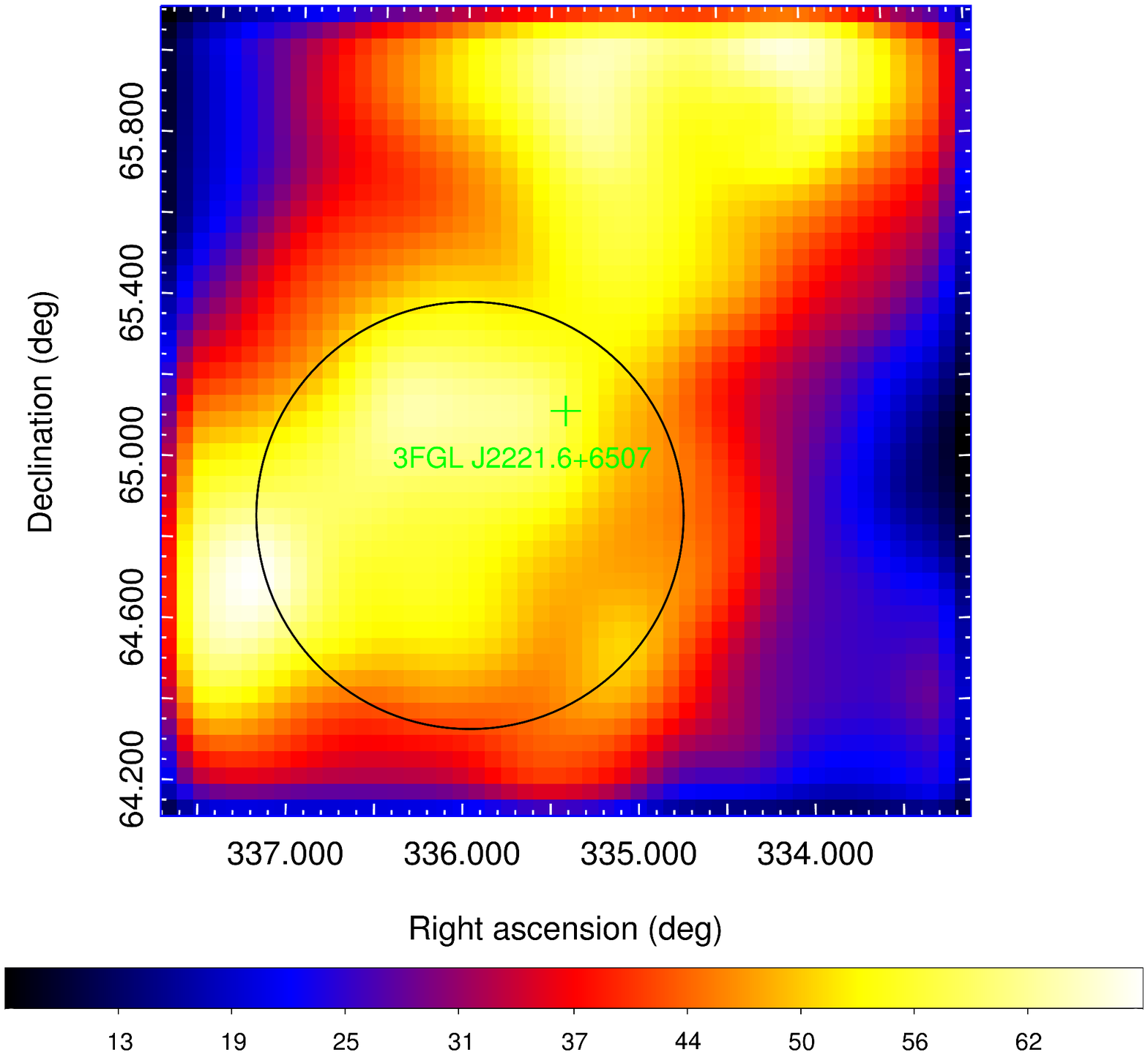}
   \caption{Examples of the targets that were found to be mixed with other
unknown sources from the TS maps. 
The green plus signs indicate the position from the LAT third source catalog, 
and the solid circles indicate the 2$\sigma$ positional error 
circles we estimated for the targets.}
   \label{fig:ts}
\end{figure}

\section{Data Analysis}
\label{sec:ar}

\subsection{\fermi\ LAT Data}
LAT, one of the two main instruments onboard \fermi, 
is an imaging $\gamma$-ray telescope conducting all-sky survey in the 
energy range from 20 MeV to 300 GeV. It was 
designed such that $\gamma$-ray events are distinguished from the  
background events through measuring the direction, energy, and arrival 
time of each $\gamma$-ray photon \citep{atw+09}. In the analysis of this paper, 
the data for each target we used are 
selected from Fermi Pass 7 Reprocessed database within 
15 deg of the target's position. The time period spans from 2008 August 4 
15:43:39 to 2015 January 22 16:08:17 (UTC; nearly 6.5 yrs), and 
the energy range is from 
200 MeV to 300 GeV to avoid the relative large uncertainties of the instrument 
response function of the LAT in the low energy range. Following the 
recommendations of the LAT team, we selected events with zenith angles 
less than 100 deg to exclude possible contamination from the Earth's limb.  

\subsubsection{Maximum Likelihood Analysis}

For each of our targets, we performed a standard binned maximum 
likelihood analysis \citep{mat+96} to the data using the LAT science tools 
software package v9r33p0. Based on the LAT third source catalog, all sources 
within 25 deg centered at the position of each target 
were included to make the source model. The spectral parameters 
of these sources are provided in the catalog.  
The spectral normalization parameters of the sources within 5 deg from 
each target, which were considered if they were detected with 
$>5\sigma$ significance, were set free. All the other 
parameters were fixed at their catalog values. Considering the Galactic and the 
extragalactic diffuse emission, we added the model gll\_iem\_v05\_rev1.fits 
and the spectrum file iso\_source\_v05.txt to the source model. 
The normalization parameters of the diffuse emission were left free as well.
\begin{figure}
   \centering
   \includegraphics[width=0.36\textwidth]{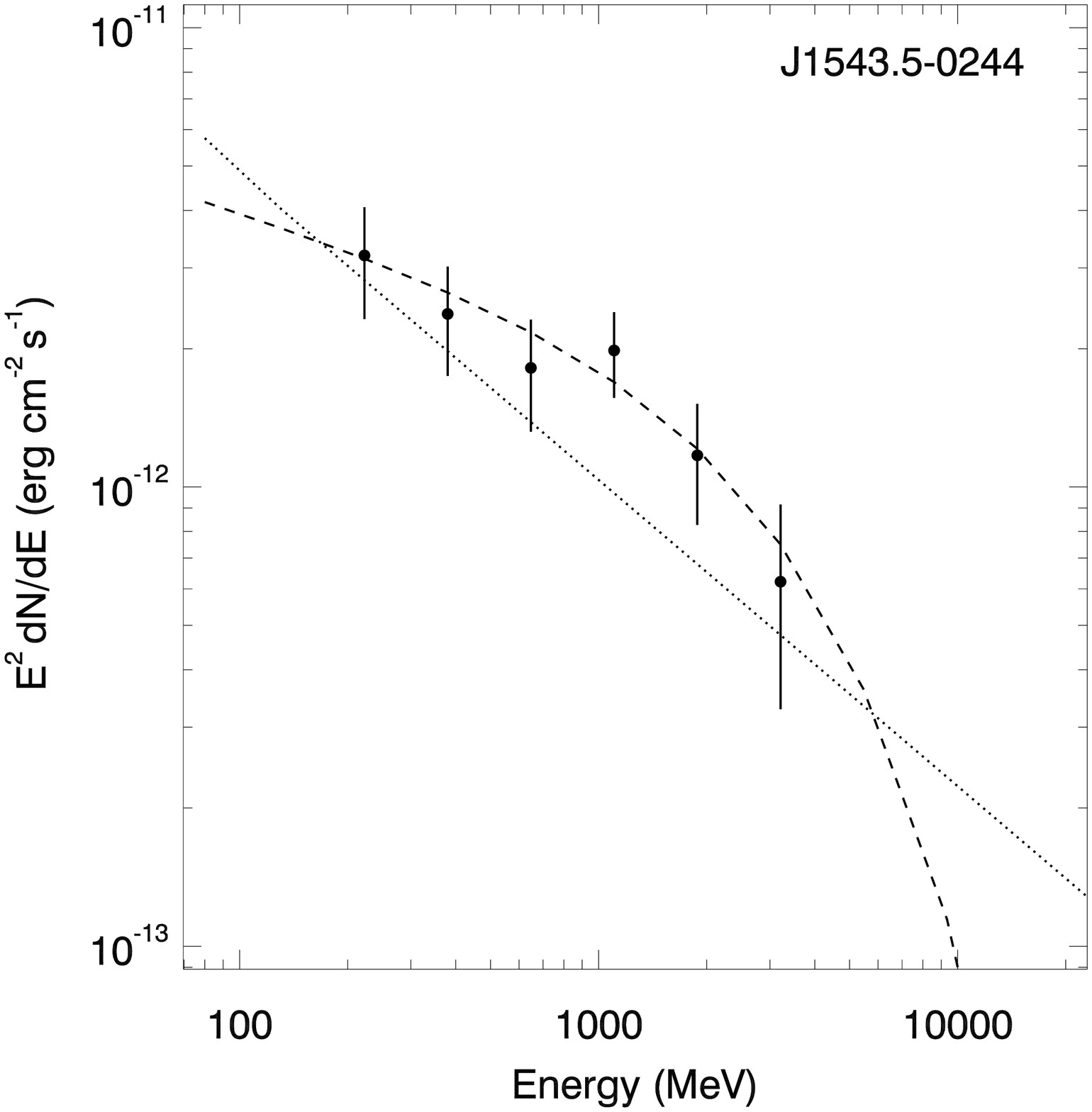}
   \includegraphics[width=0.36\textwidth]{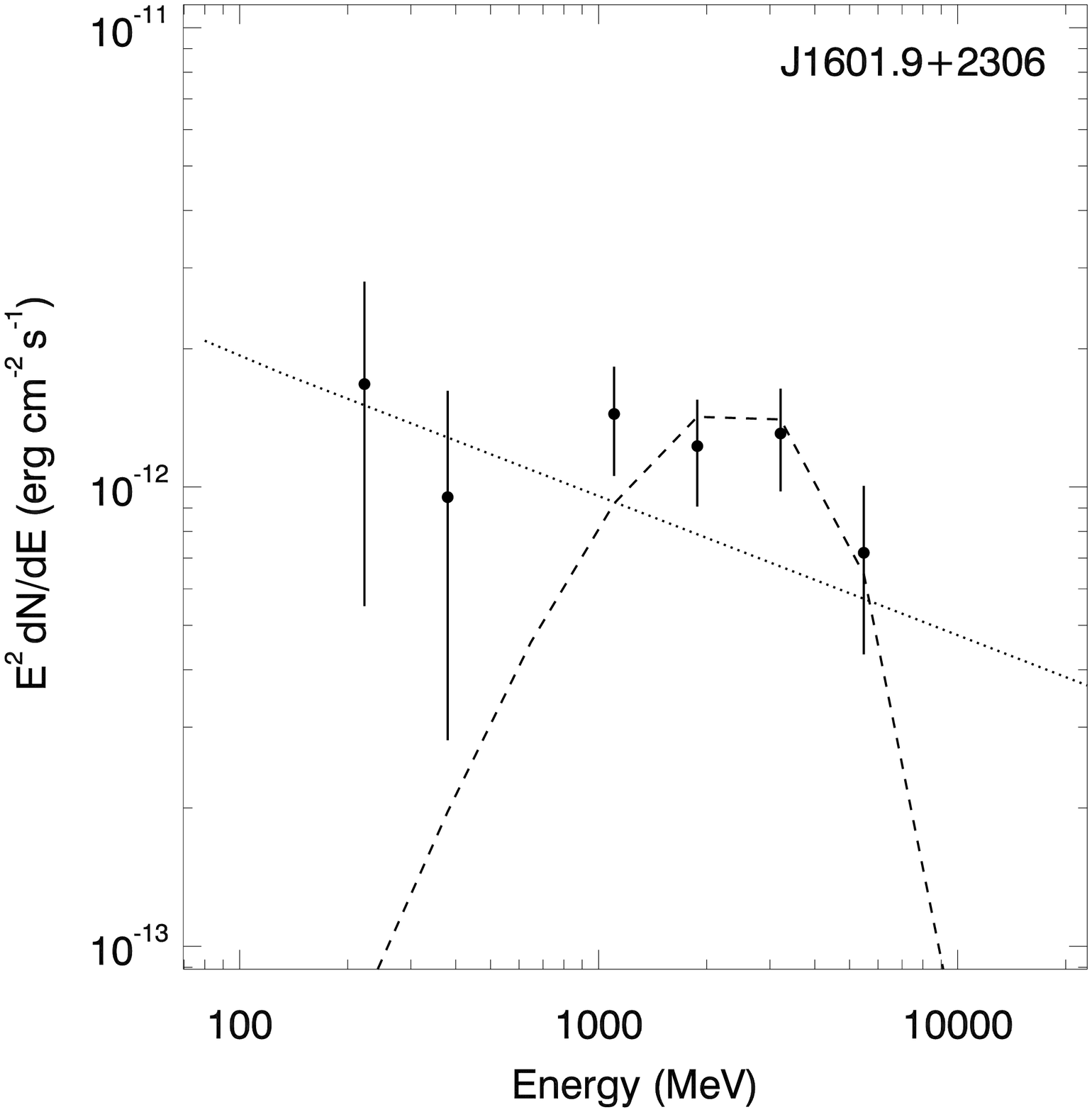}
   \includegraphics[width=0.36\textwidth]{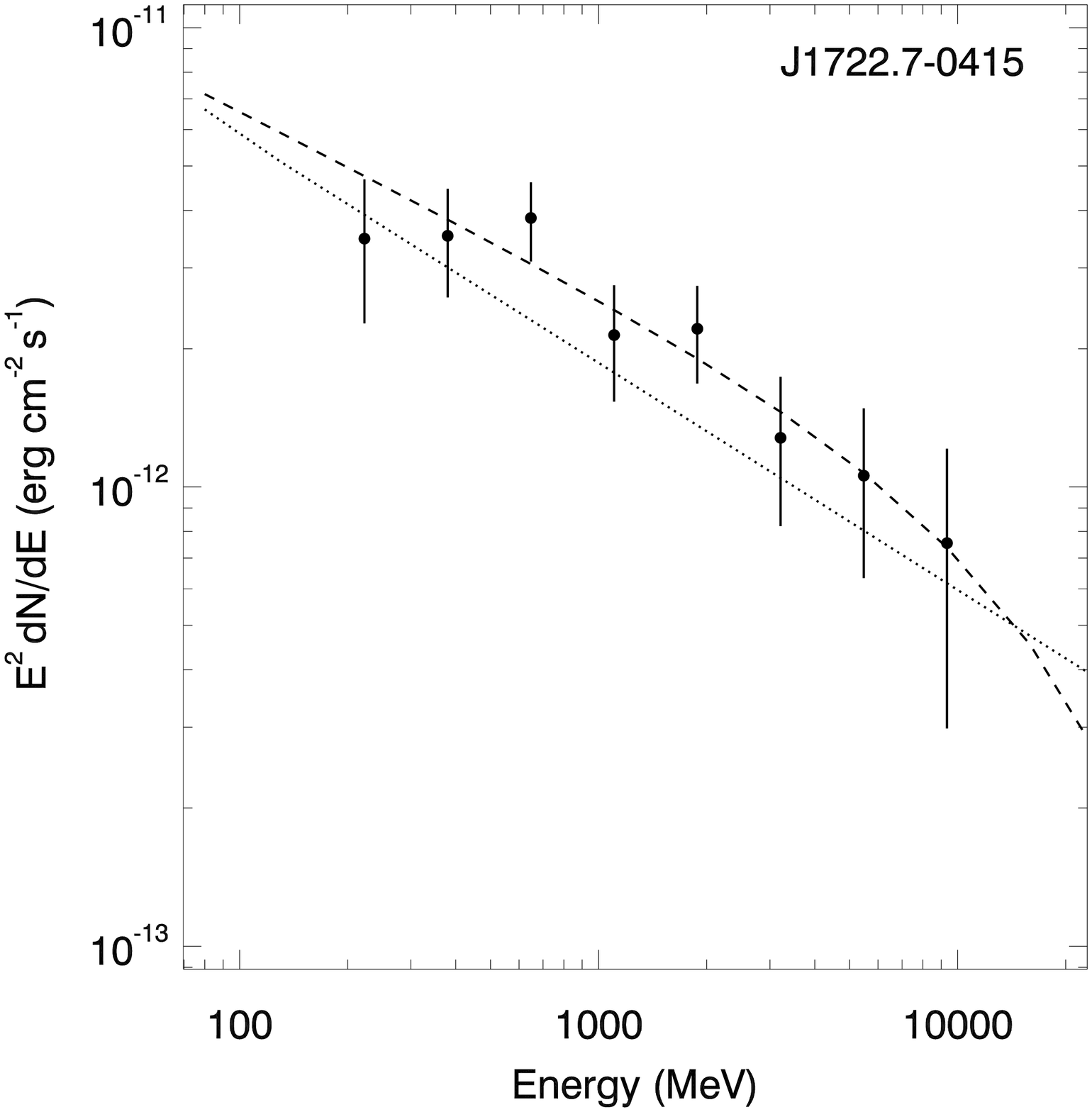}
   \caption{$\gamma$-ray spectra of J1543.5-0244, J1601.9+2306, 
and J1722.7-0415. The dotted and dashed curves are the best-fit PL and
PLE models respectively. }
   \label{fig:pls}
\end{figure}

We obtained the Test Statistic (TS) map of a 2$^\circ\times 2^\circ$ region 
centered at the position of each target. Defined as 
TS$=-2log(L_{0}/L_{1})$, where $ L_{0} $ and $ L_{1} $ respectively 
are the maximum likelihood values
for a model without and with an additional source at a specified 
location \citep{1fgl10}, the square root of a TS value is 
approximately equal to the detection significance for a given source. 
By examining the TS map of each target, we identified `isolated' point-like
sources among them, which we defined not to be mixed with other unknown sources 
or located in a region with strong, extended emission. We considered them 
as `clean' targets. 
We then ran \textit{gtfindsrc} in the LAT software package to 
determine the accurate positions for these clean targets. 
Among the initially selected 39 sources, there are 27 such clean sources.
They are listed in Tables~\ref{tab:cmsps} \& \ref{tab:nosc}.
The best-fit positions we obtained are consistent with
those provided in the LAT third source catalog within $ 2\sigma $ error circles.

The other 12 sources are not clean point sources, as indicated by the TS maps
we obtained.  In Figure~\ref{fig:ts}, two such examples are shown.
They were found to be mixed with other unknown sources and/or located in
a region with strong background. Further analysis of the groups of sources or 
the extended emission, which will help determine their true emission features, 
requires large amount of computing time. 
Therefore these 12 sources were excluded from our target list.
To be complete, their spectral parameters provided in the LAT catalog are
given here in Table~\ref{tab:bgs}. The spectra 
of J0004.2+6757 and J1827.7+1141 were fitted with a PL model, and the other 
10 sources have spectra modeled with a LogParabola model, 
\begin{equation}
\frac{dN}{dE} = N_0\left(\frac{E}{E_b}\right)^{-\alpha-\beta log \left(E/E_b\right)},
\end{equation}
where $N_{0}$ , $ \alpha $, and $ \beta $ are flux density, photon index, 
and the curvature, respectively. The energy $ E_b $ was set such that errors
on differential fluxes were minimal, and “Signif\_curve” is the curvature 
significance estimated from likelihood 
values for a PL model or a LogParabola model.
\begin{figure}
   \centering
   \includegraphics[width=0.36\textwidth]{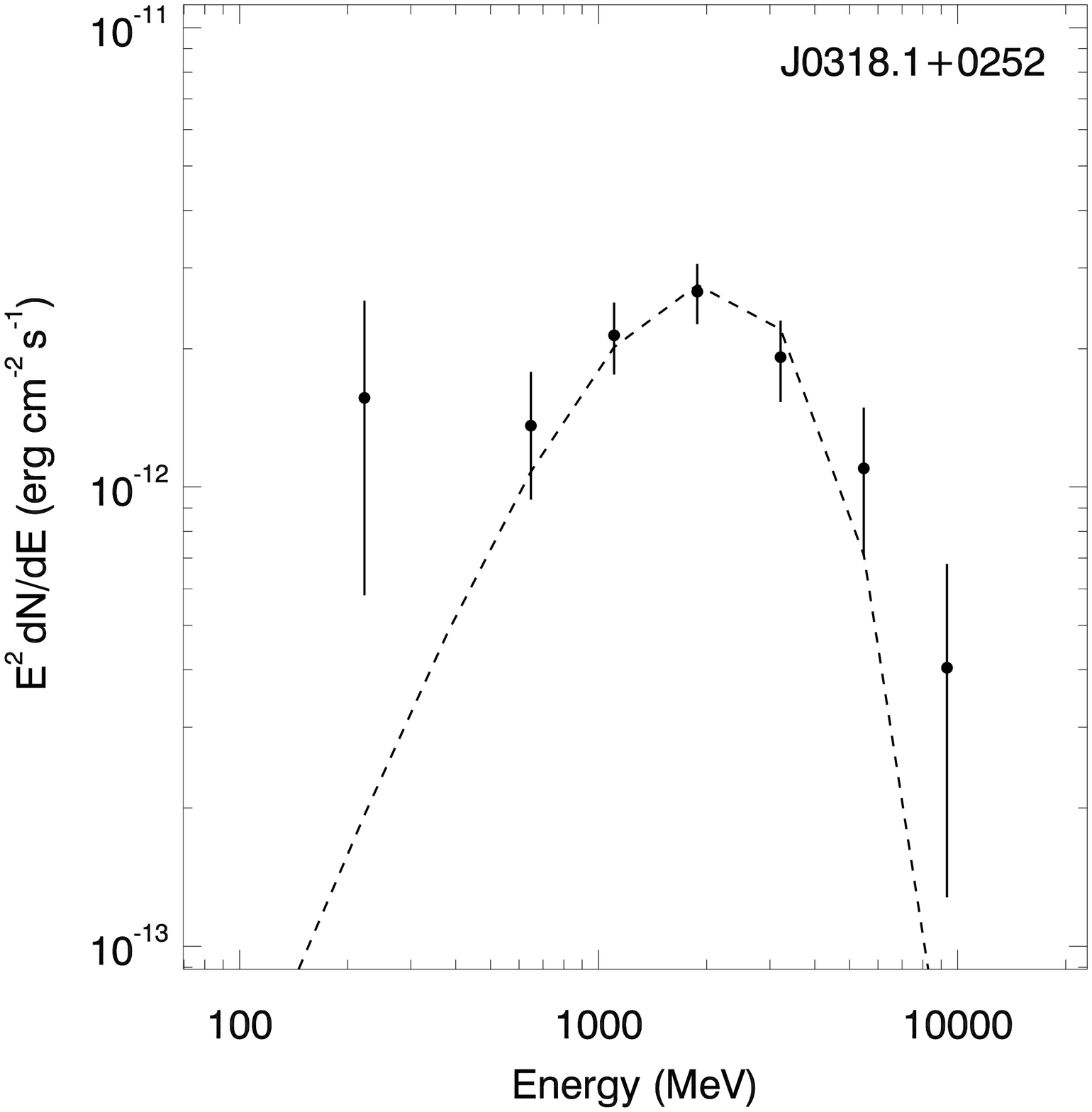}
   \includegraphics[width=0.36\textwidth]{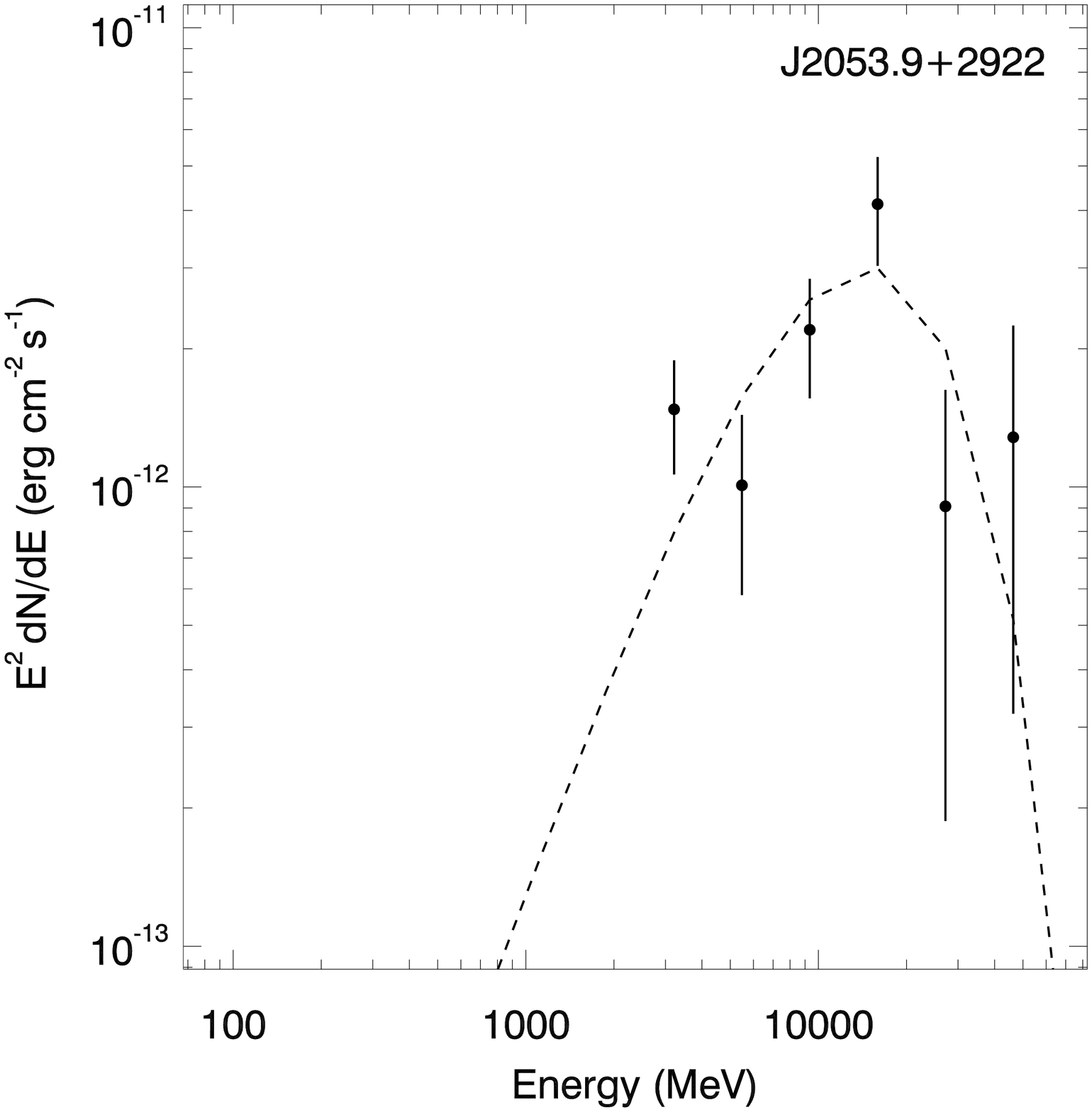}
   \caption{$\gamma$-ray spectra of J0318.1+0252 and J2053.9+2922, with
the dashed curves being the best-fit PLE models. }
   \label{fig:nmsps}
\end{figure}

\subsubsection{Spectral Analysis}
\label{subsec:sa}

We extracted the $\gamma$-ray spectra for the clean point-like sources 
by performing two separate fits at their best-fit positions.
First, We modeled each source with a simple power law (PL)
\begin{equation}
\frac{dN}{dE} = N_0 \left(\frac{E}{E_0}\right)^{-\Gamma},
\end{equation}
where $N_{0}$ is the normalization, $\Gamma$ is the photon index, and 
we set $E_{0} = 1$ GeV.  We evenly divided energy logarithmically from 0.1 to 
300 GeV into 15 energy bands for the spectra analysis,
and kept the photon index 
fixed to the value obtained from running \textit{gtlike} at the best-fit 
position. 
For our results, only spectral data points with $ TS >4 $ were kept. 
As mentioned above, pulsars generally have exponentially cutoff power-law 
spectra.  We secondly repeated the analysis using a power law with 
an exponential cutoff (PLE)
\begin{equation}
\frac{dN}{dE} = N_0 \left(\frac{E}{E_0}\right)^{-\Gamma}\exp(-\frac{E}{E_c}),
\end{equation}
where $E_{c}$ is the cutoff energy. By comparing results from the two spectral
models, the curvature significance $Signif\_curve$ was obtained, which
was estimated from $Signif\_curve=\sqrt{2log(L_{PLE}/L_{PL})}$,
where $ L_{PLE} $ and $ L_{PL} $ are the maximum likelihood values modeled
with PLE and PL, respectively.

From this analysis, we found that for three sources, whose spectral
results are given in Table~\ref{tab:nosc},
a PLE model is not significantly better than a PL one. 
Among them, J1601.9+2306 had a TS value
from a PL model larger than that from a PLE.
Their spectra are shown in Figure~\ref{fig:nmsps}.
We therefore excluded these three sources from our target list.

\subsubsection{Variability Analysis}

As a further check, we performed timing analysis of the LAT data for 
the 24 remaining sources. The time period from 2008 August 4 
23:59:59 to 2014 December 31 23:59:57(UTC) was divided into 30-day intervals.
We adopted the power law model 
leaving the photon index fixed at the value obtained in Section~\ref{subsec:sa}
and conducted likelihood analysis in each time bin at the best-fit position 
of each source. 
The light curves and TS curves were thus extracted.
No significant flux variations were seen from the 30-day interval light curves,
which are consistent with the results in the LAT third source catalog for them.
\begin{figure}
   \centering
   \includegraphics[width=0.36\textwidth]{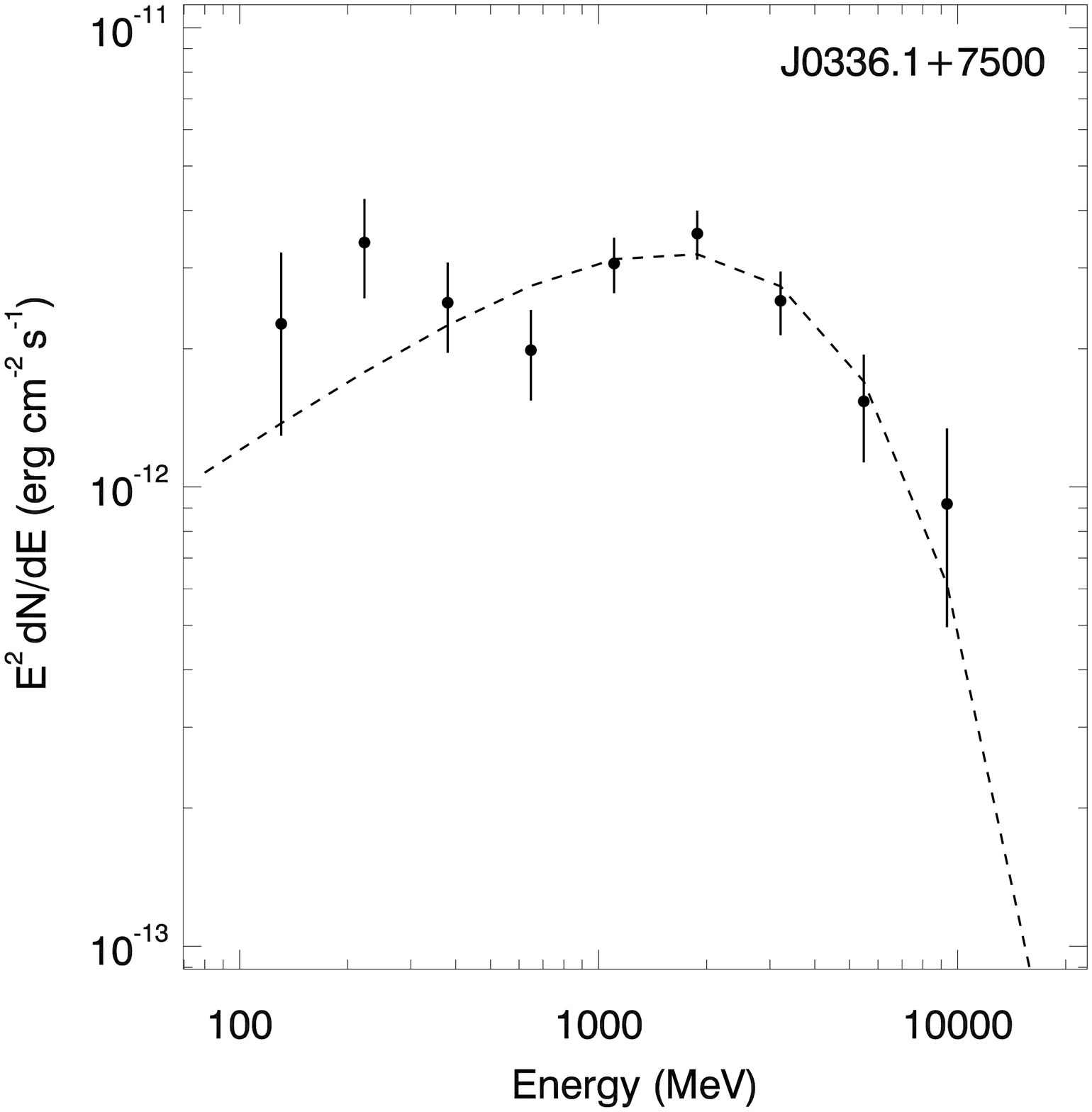}
   \includegraphics[width=0.36\textwidth]{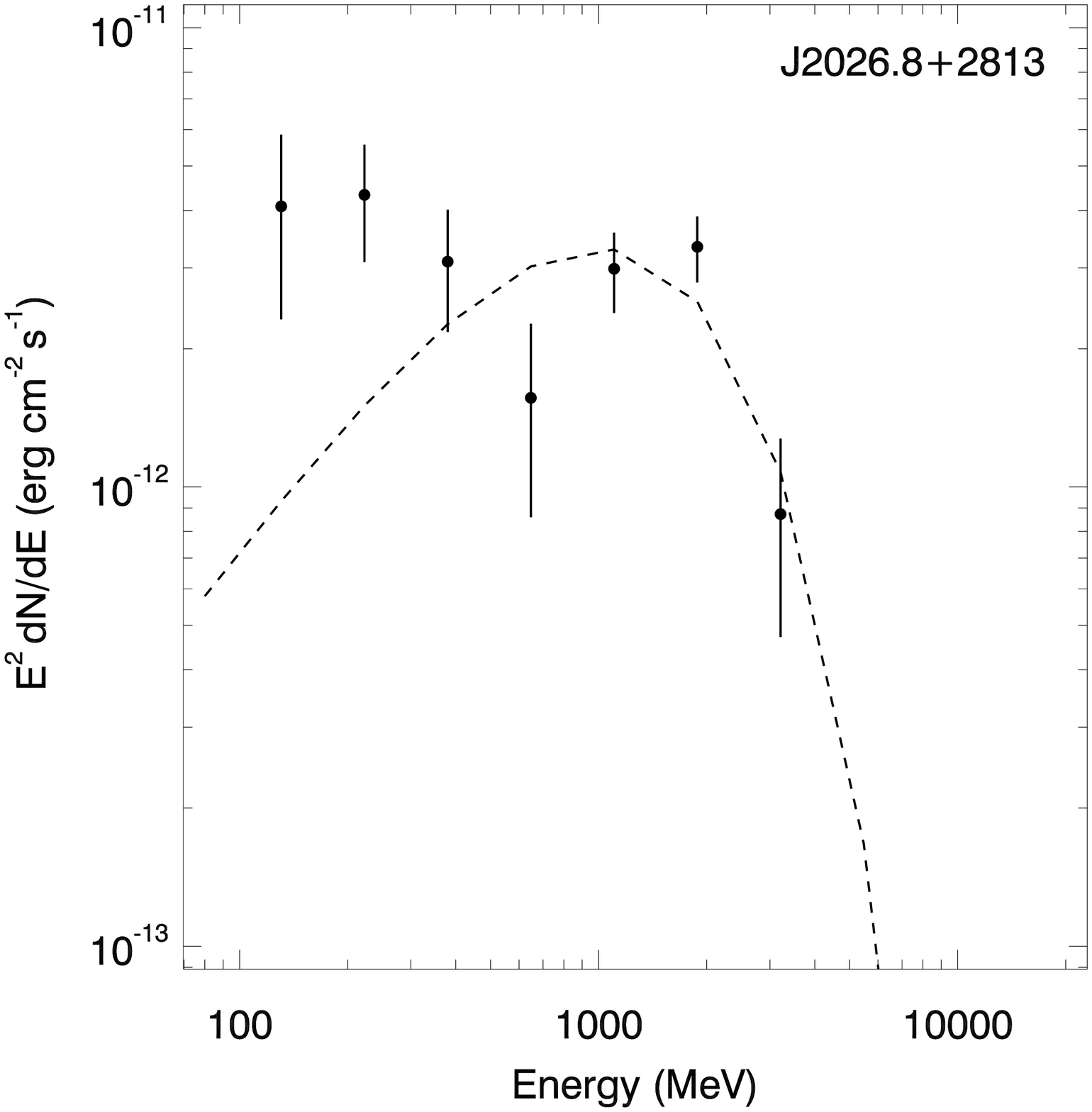}
   \caption{$\gamma$-ray spectra of J0336.1+7500 and J2026.8+2813.
The dashed curves indicate the best-fit PLE models, which do not well
describe the spectral data points as the spectra probably have two components.
    }
   \label{fig:tcs}
\end{figure}

\subsection{X-ray data analysis}

We searched for possible X-ray counterparts to the 24 $\gamma$-ray sources.
Among them, J1544.6$-$1125 has been studied at multiple wavelengths (including
X-rays) and suggested to be a transitional MSP due to its similar emissional 
properties \citep{bh15,bog15}, and J1625.1$-$0021 and J1653.6$-$0158 
have
been studied at X-ray energies and searched for potential optical/infrared 
counterparts \citep{kon+14,hui+15}. No analysis of their archival data was
conducted.
For each of the rest of the targets, its $2\sigma$ positional uncertainty region
covered by any archival X-ray imaging data from {\it XMM-Newton}, 
{\it Chandra}, or {\it Swift} telescopes were searched and analyzed.
We found that four $\gamma$-ray sources had X-ray sources in their 
uncertainty regions, 16 were covered by short \textit{Swift}
observations but with no detection of any X-ray sources. 

For the detections, the {\it Chandra} data were reprocessed using 
the script {\it chandra\_repro} in the {\it Chandra} Interactive Analysis 
Observation software ({\sc ciao 4.6}). We used the source detection tool 
in {\sc ciao} ({\sc celldetect}) for source detection.
A 10$''$ radius cicular region centered at a source was used to extract 
the source's photons, and a nearby source-free region with the same size 
was taken as the background.
The source and background spectra were obtained with the {\sc ciao} tool
{\sc psextract.}
We used $\chi^2$ statistics in the spectral fitting.
\begin{figure}
   \centering
   \includegraphics[width=0.36\textwidth]{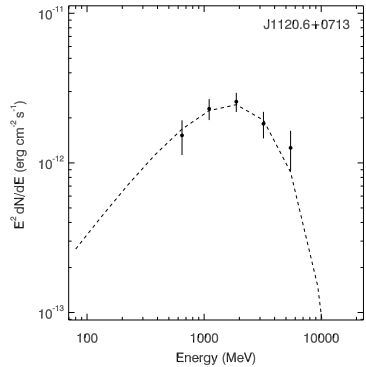}
   \caption{$\gamma$-ray spectrum of J1120.6+0713, with the dashed curve
indicating the best-fit PLE model.}
   \label{fig:nagns}
\end{figure}

Among the available \textit{Swift} data for each target, 
we selected the dataset with the longest exposure time when there are
multiple sets of data.
The {\it Swift} XRT data were processed using the XRTDAS 
software included in the HEASOFT package (version 6.13) distributed by 
the High Energy Astrophysics Science Archive Research Center (HEASARC). 
For each observation, calibrated and cleaned PC-mode event 
files were produced with the {\sc xrtpipeline} task. We used 
the {\sc ximage} detection algorithm, {\sc detect}, to locate X-ray 
point sources in the XRT images. 
The positions of the detected sources were then refined with using
the task {\sc xrtcentroid} of the XRTDAS package. 
We extracted photons from a circular region with 47\arcsec\ radius 
around a source and from a nearby source-free region with the same size as 
the background.  We adopted the Cash Statistic \citep{cas79} for spectral 
modelling due to the few net counts.

In the spectral fitting, due to the limited photon counts for
most of our sources, only an absorbed power law was used as the model, where
we fixed the absorption column density to the Galactic value \citep{dl90} in 
the direction of each source.  The obtained spectral parameters are given 
in Table~\ref{tab:xray}.  The power law photon indices range from $\sim 1 - 2$ 
for these sources and suggest that the X-ray emission is mostly non-thermal in 
nature. However, for the source J1627.8+3217, the power law model results in 
a large photon index, $\Gamma \sim 3.1$, which probably suggests a 
thermal scenario instead for this source. We thus also examined its 
spectrum with an absorbed blackbody model, where the absorption was 
fixed at the Galactic value. We found a temperature 
of $kT = 0.19^{+0.11}_{-0.06}\rm~keV$ with C = 5.9 for 3 degree of
freedom (DoF). This model 
provides a more reasonable description of the data.
For the source J2103.7-1113, two X-ray sources within 
the $2\sigma$ {\it Fermi} error circle were detected, and both were 
well described by an absorbed power law model.

For the non-detections, which resulted from short \textit{Swift} observations,
we estimated 3$\sigma$ upper 
limits on fluxes from the count rates using the 
webPIMMS\footnote{http://heasarc.gsfc.nasa.gov/cgi-bin/Tools/w3pimms/w3pimms.pl}.
An absorbed power law spectra with 1.7 photon index was assumed, and
the absorption column density to a source was fixed at the Galactic value in 
the direction of the source \citep{dl90}. The flux upper limits for 
the \fermi\ sources are given in the Table~\ref{tab:noxray}.   

\section{Results and Discussion}
\label{sec:rd}

Having analyzed the LAT data of 39 un-associated sources selected from
the 3rd LAT source catalog, we found 27 clean point-like sources among them.
Further requiring curvature significance in a spectrum,
24 sources were selected.  Their spectral results are provided 
in Table~\ref{tab:cmsps}.  Among the 24 sources, J1544.6$-$1125 has been 
already well studied at 
multiple wavelengths, particularly X-rays, and suggested to be
a transitional MSP \citep{bh15,bog15}. The sources
J1625.1$-$0021 and J1653.6$-$0158 have been studied as well
and are listed as promising candidate MSPs \citep{hui+15}. Moreover, an 
orbital period of 75 min was found for the second source from optical imaging, 
indicating its likely nature of being an MSP binary \citep{kon+14}.
These studies support our target selection and further data
analysis selection.

Examining the obtained spectra, we note that since MSPs generally 
have \gr\ spectra
with $\Gamma$ and $E_c$ in the ranges of 0.4--2.0 and 1.1--5.4\,GeV respectively
(see \citealt{2fpsr13} for details), the sources J0318.1$+$0252 and 
J2053.9$+$2922 have the parameters of $\Gamma=0.00\pm0.03$ and 0.2$\pm$0.5,
and $E_c=1.0\pm0.3$\,GeV and $8\pm3$\,GeV, respectively. The values, 
particularly for the first source, are not within the ranges defined from 
the known \gr\ MSPs.
Their spectra are shown in Figure~\ref{fig:nmsps}. As can be seen, the
spectra have a fast drop in the low, 0.1--1 GeV energy range, not 
containing significant emission and thus making $\Gamma$ close to zero. 
Given the spectral property, they likely are not MSPs.  
In addition, the sources J0336.1+7500 and J2026.8+2813 appear to possibly
have two components in their spectra, which are shown in Figure~\ref{fig:tcs}.
We examined their TS maps and they are consistent with being point sources.
Their TS maps at low (0.1--1.0\,GeV) and high (1.0-300\,GeV) energy ranges 
were also calculated, but no evidence was found from the TS maps for 
the cases such as the presence of an additional source in the region.
These two sources are of interest for further investigation.

We note that source J1120.6+0713 was listed as an AGN
in the first catalog of AGN detected by \textit{Fermi} LAT (associated with
CRATES J1120+0704; \citealt{abd+10}), but it was not in the third catalog
anymore \citep{ack+15}. The spectrum we obtained, which is shown in
Figure~\ref{fig:nagns}, is well described by a PLE model. In addition,
six sources in Table~\ref{tab:cmsps} were listed as promising
dark matter subhalo candidates in \citet{bhl15}. The double identification is
due to their selection criteria of non-variables with $>$20$^{\circ}$ 
Galactic latitudes (similar to ours) and curved spectra calculated
from dark matter annihilation models. In any case,  
the likely MSPs J1544.6$-$1125 and J1625.1$-$0021 (see above) 
are in their list too, indicating the possibly high chance of identifying 
an MSP as a candidate dark matter subhalo. Information about possible
nature of these sources is provided in Table~\ref{tab:cmsps}.

Finally for the candidate MSPs in Table~\ref{tab:cmsps} that were covered
by X-ray observations, we calculated their $G_{100}$ flux, which
is defined as the total \gr\ flux in the energy range of
0.1--100\,GeV \citep{2fpsr13}. The \gr--to--X-ray flux ratios (for the cases
of having X-ray sources in the source field) or low limits
on the flux ratios (for the cases of non-detection) were then estimated.
The values are given in Table~\ref{tab:xray} and \ref{tab:noxray}. For most
of the known \gr\ MSPs, the ratios are in a range of 100--1000 (see Table~16
in \citealt{2fpsr13}). This property suggests that none of the X-ray sources
listed in Table~\ref{tab:xray} are the counterparts. In addition, if we
consider that the sources in Table~\ref{tab:noxray} are MSPs, their low
flux-ratio limits
of $>$20--100 suggest that the X-ray observations are not sufficiently deep
for detecting any X-ray counterparts. Further X-ray observations of them
are needed in order to identify their MSP nature by finding X-ray 
counterparts. 

\begin{acknowledgements}
This research made use of the High Performance Computing Resource in the Core
Facility for Advanced Research Computing at Shanghai Astronomical Observatory.
This research was supported by the National Natural Science Foundation
of China (11373055) and the Strategic Priority Research Program
``The Emergence of Cosmological Structures" of the Chinese Academy
of Sciences (Grant No. XDB09000000). Z.W. acknowledges the support 
by the CAS/SAFEA International Partnership Program for Creative Research Teams;
J.V. by the Chinese Academy of Sciences President’s International 
Fellowship Initiative (CAS PIFI, Grant No. 2015PM059); and
Y.X. by the Shanghai Natural Science 
Foundation for Youth (13ZR1464400) and the National Natural Science Foundation
of China for Youth (11403075). 
\end{acknowledgements}

\begin{table}
\bc
\begin{minipage}[]{100mm}
  \caption[]{Spectral results for candidate millisecond pulsars.\label{tab:cmsps}}
\end{minipage}
\small
  \begin{tabular}{llcccccc}
\hline
Source name  & Spectra model    & Flux/$10^{-9}$    & $\Gamma$        & E$_{c}$        & TS    & Signif\_Curve & Comments \\
\qquad       & \qquad  & (photons cm$^{-2}$\,s$^{-1}$) & \qquad          & (GeV)          &\qquad & ($\sigma$) \\
\hline
J0212.1+5320 & PowerLaw         & 14.5 $\pm$ 0.9       & 2.17 $\pm$ 0.04 & \qquad         & 848   & 9.11 & \\
\qquad       & PLSuperExpCutoff & 10.8 $\pm$ 0.9       & 1.3 $\pm$ 0.1   & 2.6 $\pm$ 0.5  & 924   & \qquad & \\	
\hline
J0238.0+5237 & PowerLaw         & 12 $\pm$ 1           & 2.38 $\pm$ 0.06 & \qquad         & 319   & 5.14  & \\
\qquad       & PLSuperExpCutoff & 10 $\pm$ 1           & 1.8 $\pm$ 0.2   & 4 $\pm$ 1      & 341   & \qquad & \\	
\hline
J0312.1$-$0921 & PowerLaw         & 6.0 $\pm$ 0.8        & 2.26 $\pm$ 0.08 & \qquad         & 190   & 5.13 & c-subhalo \\
\qquad       & PLSuperExpCutoff & 4.0 $\pm$ 0.8        & 1.2 $\pm$ 0.3   & 2.0 $\pm$ 0.6  & 211   & \qquad & \\	
\hline
J0318.1+0252 & PowerLaw         & 5.8 $\pm$ 0.7        & 2.19 $\pm$ 0.07 & \qquad         & 191   & 6.54 & c-subhalo \\
\qquad       & PLSuperExpCutoff & 2.6 $\pm$ 0.4        & 0.00 $\pm$ 0.03 & 1.0 $\pm$ 0.3  & 231   & \qquad & non-MSP \\	
\hline
J0336.1+7500 & PowerLaw         & 9.5 $\pm$ 0.8        & 2.24 $\pm$ 0.05 & \qquad         & 389   & 6.67 & \\
\qquad       & PLSuperExpCutoff & 7.1 $\pm$ 0.8        & 1.5 $\pm$ 0.2   & 3.0 $\pm$ 0.8  & 431   & \qquad & \\
\hline
J0545.6+6019 & PowerLaw         & 5.6 $\pm$ 0.7        & 2.03 $\pm$ 0.06 & \qquad         & 279   & 5.15 & \\
\qquad       & PLSuperExpCutoff & 4.0 $\pm$ 0.6        & 1.4 $\pm$ 0.2   & 7 $\pm$ 2      & 303   & \qquad & \\
\hline
J0758.6$-$1451 & PowerLaw         & 7.5 $\pm$ 0.9        & 2.32 $\pm$ 0.07 & \qquad         & 212   & 4.91 & \\
\qquad       & PLSuperExpCutoff & 5.2 $\pm$ 0.9        & 1.4 $\pm$ 0.3   & 2.3 $\pm$ 0.8  & 234   & \qquad & \\
\hline
J0935.2+0903 & PowerLaw         & 6.2 $\pm$ 0.8        & 2.5 $\pm$ 0.1   & \qquad         & 135   & 3.32 & \\
\qquad       & PLSuperExpCutoff & 5.0 $\pm$ 0.9        & 1.7 $\pm$ 0.4   & 1.8 $\pm$ 0.8  & 145   & \qquad & \\
\hline
J0953.7$-$1510 & PowerLaw         & 5.4 $\pm$ 0.6        & 2.13 $\pm$ 0.07 & \qquad         & 227   & 6.73 & c-subhalo \\
\qquad       & PLSuperExpCutoff & 2.1 $\pm$ 0.6        & 0.6 $\pm$ 0.4   & 1.5 $\pm$ 0.4  & 269   & \qquad & \\	
\hline
J1120.6+0713 & PowerLaw         & 6.0 $\pm$ 0.7        & 2.20 $\pm$ 0.07 & \qquad         & 249   & 6.49 & AGN (?) \\
\qquad       & PLSuperExpCutoff & 4.0 $\pm$ 0.6        & 1.0 $\pm$ 0.3   & 1.7 $\pm$ 0.4  & 292   & \qquad & \\
\hline  
J1225.9+2953 & PowerLaw         & 7.0 $\pm$ 0.7        & 2.11 $\pm$ 0.06 & \qquad         & 436   & 6.59 & c-subhalo \\
\qquad       & PLSuperExpCutoff & 4.7 $\pm$ 0.7        & 1.3 $\pm$ 0.2   & 3.3 $\pm$ 0.8  & 469   & \qquad & \\	
\hline
J1544.6$-$1125 & PowerLaw         &  12 $\pm$ 1          & 2.54 $\pm$ 0.07 & \qquad         & 262   & 3.47 & c-MSP \\
\qquad       & PLSuperExpCutoff &  11 $\pm$ 1          & 2.1 $\pm$ 0.2   & 3 $\pm$ 2      & 27    & \qquad & \\	
\hline
J1625.1$-$0021 & PowerLaw         & 16.5 $\pm$ 0.8       & 2.09 $\pm$ 0.03 & \qquad         & 1261  & 13.16 & c-MSP \\
\qquad       & PLSuperExpCutoff & 10.5 $\pm$ 0.8       & 0.8 $\pm$ 0.2   & 1.9 $\pm$ 0.2  & 1433  & \qquad & \\	
\hline
J1627.8+3217 & PowerLaw         & 3.6 $\pm$ 0.5        & 2.15 $\pm$ 0.08 & \qquad         & 158   & 4.55 & \\
\qquad       & PLSuperExpCutoff & 2.4 $\pm$ 0.5        & 1.2 $\pm$ 0.3   & 3 $\pm$ 1      & 178   & \qquad & \\	
\hline
J1653.6$-$0158 & PowerLaw         &  31 $\pm$ 1          & 2.32 $\pm$ 0.03 & \qquad         & 1686  & 9.43 & c-MSP \\
\qquad       & PLSuperExpCutoff &  27 $\pm$ 1          & 1.75 $\pm$ 0.08 & 3.3 $\pm$ 0.5  & 1747  & \qquad & \\	
\hline
J1730.6$-0$357 & PowerLaw         &   7 $\pm$ 1          & 2.17 $\pm$ 0.08 & \qquad         & 124   & 4.49 & \\
\qquad       & PLSuperExpCutoff &   4 $\pm$ 1          & 1.1 $\pm$ 0.4   & 3 $\pm$ 1      & 143   & \qquad & \\
\hline
J1950.2+1215 & PowerLaw         &  15 $\pm$ 2          & 2.9 $\pm$ 0.1   & \qquad         & 149   & 3.13 & \\
\qquad       & PLSuperExpCutoff &  13 $\pm$ 2          & 2.2 $\pm$ 0.3   & 2 $\pm$ 1      & 151   & \qquad & \\
\hline		
J2026.8+2813 & PowerLaw         &  10 $\pm$ 1          & 2.57 $\pm$ 0.09 & \qquad         & 87    & 4.91 & \\
\qquad       & PLSuperExpCutoff &   7 $\pm$ 2          & 0.9 $\pm$ 0.6   & 0.9 $\pm$ 0.3  & 111   & \qquad & \\ 
\hline
J2053.9+2922 & PowerLaw         & 1.3 $\pm$ 0.3        & 1.59 $\pm$ 0.09 & \qquad         & 114   & 5.52 & non-MSP\\
\qquad       & PLSuperExpCutoff & 0.5 $\pm$ 0.1        & 0.2 $\pm$ 0.5   & 8 $\pm$ 3      & 146   & \qquad & \\	
\hline 
J2103.7$-$1113 & PowerLaw         & 6.2 $\pm$ 0.7        & 2.18 $\pm$ 0.07 & \qquad         & 239   & 4.31 & c-subhalo \\
\qquad       & PLSuperExpCutoff & 4.4 $\pm$ 0.8        & 1.4 $\pm$ 0.3   & 3 $\pm$ 1      & 255   & \qquad & \\	
\hline
J2117.6+3725 & PowerLaw         &  15 $\pm$ 1          & 2.57 $\pm$ 0.06 & \qquad         & 314   & 4.16 & \\
\qquad       & PLSuperExpCutoff &  14 $\pm$ 1          & 2.0 $\pm$ 0.2   & 2.5 $\pm$ 0.9  & 327   & \qquad & \\	
\hline
J2212.5+0703 & PowerLaw         & 7.1 $\pm$ 0.8        & 2.27 $\pm$ 0.07 & \qquad         & 209   & 5.29 & c-subhalo \\
\qquad       & PLSuperExpCutoff & 5.0 $\pm$ 0.8        & 1.4 $\pm$ 0.3   & 2.5 $\pm$ 0.8  & 236   & \qquad & \\
\hline
J2233.1+6542 & PowerLaw         &  21 $\pm$ 2          & 2.69 $\pm$ 0.07 & \qquad         & 240   & 4.60 & \\
\qquad       & PLSuperExpCutoff &  19 $\pm$ 2          & 1.9 $\pm$ 0.3   & 1.6 $\pm$ 0.6  & 252   & \qquad & \\ 
\hline
J2250.6+3308 & PowerLaw         & 5.0 $\pm$ 0.8        & 2.5 $\pm$ 0.1   & \qquad         & 81    & 3.97 & \\
\qquad       & PLSuperExpCutoff & 4.0 $\pm$ 0.8        & 1.2 $\pm$ 0.5   & 1.1 $\pm$ 0.5  & 96    & \qquad & \\
\hline
  \end{tabular}
\ec
\end{table}
\begin{table} 
\bc
\begin{minipage}[]{100mm}
\caption[]{Sources without sufficient curvature significance.
\label{tab:nosc}}
\end{minipage}
\small
\begin{tabular}{llccccc}
\hline
Source name  & Spectra model    & Flux/$10^{-9}$                 & $\Gamma$       & E$_{c}$      & TS     & Signif\_Curve \\
\qquad       &\qquad   & (photon\,cm$^{-2}$\,s$^{-1}$) & \qquad         & (GeV)        & \qquad & ($\sigma$)    \\
\hline
J1543.5$-$0244 & PowerLaw         & 8$\pm$ 1             & 2.7$\pm$ 0.1   & \qquad       & 103    & 2.22          \\
\qquad       & PLSuperExpCutoff & 7$\pm$ 1             & 2.2$\pm$ 0.3   & 4$\pm$ 3     & 107    & \qquad        \\ 
\hline
J1601.9+2306 & PowerLaw         & 4.4$\pm$ 0.8         & 2.3$\pm$ 0.1    & \qquad         & 107   & 3.55          \\
\qquad       & PLSuperExpCutoff & 1$\pm$ 2             & 0.0$\pm$ 0.1    & 1.2$\pm$ 0.6   & 102   & \qquad        \\ \hline
J1722.7$-$0415 & PowerLaw         & 11$\pm$ 1            & 2.49$\pm$ 0.09 & \qquad       & 121    & 0.68          \\
\qquad       & PLSuperExpCutoff & 11$\pm$ 2            & 2.4$\pm$ 0.2   & 24$\pm$ 34   & 121    & \qquad        \\
\hline
\end{tabular}
\ec
\end{table}

\begin{table} 
\bc
\begin{minipage}[]{100mm}
\caption[]{Sources without clean background. \label{tab:bgs}}
\end{minipage}
\small
\begin{tabular}{llcccccc}	
\hline
Source name       & Spectra model & Flux density/10$^{-12}$  & $\Gamma$ & E$_{0}$ & Signif\_Avg & Signif\_Curve & \qquad  \\
\qquad        & \qquad & (photon\,cm$^{-2}$MeV$^{-1}$s$^{-1}$) & \qquad   & (MeV)   & ($\sigma$)  & ($\sigma$)    & \qquad \\
\hline
J0004.2+6757 &   PowerLaw    & 0.6  $\pm$ 0.1                & 2.5      & 1328.36 & 6.01        & 3.91          & \qquad \\           
J1827.7+1141 &   PowerLaw    & 0.18 $\pm$ 0.03               & 2.1      & 1960.79 & 6.49        & 3.81          & \qquad \\
\hline
Source name  & Spectra model & Flux density$/10^{-12}$      & $\alpha$ & $\beta$ & E$_{b}$ & Signif\_Avg & Signif\_Curve \\
\qquad       & \qquad & (photon\,cm$^{-2}$MeV$^{-1}$s$^{-1}$) &\qquad &\qquad  & (MeV)   & ($\sigma$)  & ($\sigma$)    \\
\hline
J0008.5+6853 &  LogParabola  & 4.3  $\pm$ 0.5                & 2.4      & 0.9     & 820.26  & 8.45        &    6.49       \\  
J0345.3+3236 &  LogParabola  & 6.8  $\pm$ 0.9                & 2.4      & 1.0     & 641.45  & 7.68        &    5.00       \\
J0431.7+3503 &  LogParabola  & 2.1  $\pm$ 0.3                & 2.6      & 0.3     & 941.06  & 7.85        &    4.02       \\ 
J0539.2$-$0536 &  LogParabola  & 11   $\pm$ 1                  & 2.5      & 1.0     & 556.20  & 7.35        &    5.10       \\  
J1729.9$-$0859 &  LogParabola  & 19   $\pm$ 2                  & 2.6      & 1.0     & 445.58  & 7.18        &    5.40       \\  
J2125.8+5832 &  LogParabola  & 4.4  $\pm$ 0.6                & 2.4      & 1.0     & 774.12  & 7.47        &    4.01       \\ 
J2206.5+6451 &  LogParabola  & 4.5  $\pm$ 0.5                & 2.8      & 0.7     & 803.49  & 8.60        &    6.03       \\ 
J2221.6+6507 &  LogParabola  & 6.1  $\pm$ 0.8                & 2.6      & 0.9     & 678.66  & 7.45        &    4.80       \\ 
J2221.7+6318 &  LogParabola  & 7.0  $\pm$ 0.8                & 2.5      & 1.0     & 739.95  & 8.42        &    6.65       \\  
J2310.1$-$0557 &  LogParabola  & 0.49 $\pm$ 0.08               & 1.8      & 0.8     & 1490.34 & 8.97        &    5.35       \\ 
\hline
\end{tabular}
\ec
\end{table}
\begin{table}
\bc
\begin{minipage}[]{100mm}
\caption[]{Properties of X-ray sources detected in the error circles of
four candidate MSPs.
\label{tab:xray}}
\end{minipage}
\small
\begin{tabular}{lcccccccc}
\hline

Source & R.A. & Dec. & $N_{H}/10^{20}$ & $\Gamma$ & $F^{unabs}_{0.3-10}$ & $\chi^2/$DoF & $G_{100}$ & $G_{100}/F^{unabs}_{0.3-10}$ \\
name & (h:m:s) & (${^\circ}:':''$) & ($\rm cm^{-2}$) & &  & &  & \\

\hline

J0212.1($S$) &  02:12:10.46 & +53:21:37.62 & $17.4$ & $1.04^{+0.47}_{-0.46}$ & $1.85^{+0.78}_{-0.51}$ & 3.1/4(C) & 16.6 & 9.0 \\
J0212.1($C$) & 02:12:10.50 & +53:21:38.94 & $17.4$ & $1.35^{+0.05}_{-0.06}$ & $1.77^{+0.07}_{-0.06}$ & 115/107 & 16.6 & 9.4 \\
J1120.6($S$) & 11:20:42.54 & +07:13:12.74 & $4.24$ & $0.74^{+0.96}_{-1.07}$ & $0.73^{+1.40}_{-0.44}$ & 2.4/2(C) & 6.2 & 8.5\\ 
J1627.8($S$) & 16:27:42.85 & +32:20:58.56 & $1.87$ & $3.07^{+1.56}_{-1.08}$ & $0.43^{+0.71}_{-0.19}$ & 5.6/3(C) & 3.1 & 7.2\\
J2103.7($C$) & 21:03:49.99 & -11:13:40.62 & $4.70$ & $1.71^{+0.22}_{-0.20}$ & $0.16^{+0.021}_{-0.020}$ & 24/13 & 6.9 & 43 \\
J2103.7($C$) & 21:03:52.31 & -11:11:32.66 & $4.70$ & $1.77^{+0.25}_{-0.24}$ & $0.10^{+0.021}_{-0.017}$ & 17/21 & 6.9 & 69\\
\hline
\end{tabular}
\ec
{Notes: (1) Source Name, where $S$ or $C$ indicate the \textit{Swift} 
or \textit{Chandra} observation used in the analysis, respectively, and
the observation IDs (exposure time) in the sequence of the table are 
00041276001 (3.3 ks), 14814 (30 ks), 0003164100(4.3 ks), 
00041418001 (3.5 ks), 12381 (30 ks);
(2)--(3) Right Ascension (R.A.) and Declination (Dec.) of each X-ray source, equinox J2000.0; (4) Absorption column density; (5) Power law index; (6) Unabsorbed flux in 0.3--10 keV band (in units of 10$^{-12}$\,erg\,cm$^{-2}$\,s$^{-1}$); 
(7) The $\chi^2/$DoF value for the model, where C-statistics is indicated by C; (8) \textit{Fermi} LAT flux in the energy range of 0.1--100 GeV (in units of 10$^{-12}$\,erg\,cm$^{-2}$\,s$^{-1}$); (9) Flux ratio between $G_{100}$ and $F^{unabs}_{0.3-10}$.}
\end{table}

\begin{table}
\bc
\begin{minipage}[]{100mm}
\caption[]{X-ray flux upper limits for the candidate MSPs. \label{tab:noxray}}
\end{minipage}
\small
\begin{tabular}{lcccccc}
\hline

Source & ObsID & Exp & $N_{H}/10^{20}$ & $F^{upper}_{0.3-10}/10^{-13}$ & $G_{100}/10^{-12}$ & $G_{100}/F^{upper}_{0.3-10}$ \\
name &  & (sec) & ($\rm cm^{-2}$) & ($\rm erg~cm^{-2}~s^{-1}$) & ($\rm erg~cm^{-2}~s^{-1}$) & \\

\hline

J0238.0 & 00047142003 & 3363 &  $26.3$ & $< 2.8$ & 13 & $>46$ \\	  
J0312.1 & 00047144004 & 3738 &  $6.23$ & $< 1.2$ & 5.9 & $>49$ \\ 
J0318.1 & 00084649005 & 1817 &  $8.77$ & $< 1.9$ & 5.1 & $>27$ \\ 
J0336.1 & 00047146002 & 1672 &  $14.3$ & $< 4.3$ & 11 & $>26$ \\	 
J0545.6 & 00084664005 & 2066 &  $14.7$ & $< 3.6$ & 8.3 & $>23$ \\	 
J0758.6 & 00041341002 & 2569 &  $11.6$ & $< 1.6$ & 7.3 & $>46$ \\	 
J0935.2 & 00084964006 & 668 &  $3.53$ & $< 4.5$  & 5.7 & $>13$ \\ 
J0953.7 & 00031656001 & 3517 &  $5.29$ & $< 1.3$ & 5.6 & $>43$ \\	
J1225.9 & 00041382001 & 4005 & $1.79$ & $< 0.76$ & 8.4 &  $>110$ \\	 
J1730.6 & 00084792002 & 1672 &  $14.1$ & $< 3.1$ & 7.2 & $>23$  \\ 
J1950.2 & 00085096001 & 802 &  $1.76$ & $< 8.1$ & 52 & $>64$ \\	 
J2026.8 & 00085106001 & 2462 &  $32.9$ & $< 2.2$ & 7.9 & $>36$ \\	  
J2117.6 & 00041492001 & 3716 &   $17.0$ & $< 1.2$ & 15 & $>130$ \\	 
J2212.5 & 00047320001 & 2497 &   $6.47$ & $< 3.2$ & 7.3 & $>23$ \\ 
J2233.1 & 00084887001 & 2670 &   $59.1$ & $< 2.6$ & 20 & $>77$ \\ 
J2250.6 & 00085140002 & 1952 &   $7.87$ & $< 2.4$ & 4.5 & $>19$ \\	 
\hline
\end{tabular}
\ec
{Notes: (1) Source Name; (2) ID of the \textit{Swift} observation used for the analysis; (3) Exposure time in seconds for each observation; (4) Absorption column density; (5) The $3\sigma$ upper limit on flux in 0.3--10 keV band (a power law with 1.7 photon index was assumed); (6) \textit{Fermi} LAT flux in the energy range of 0.1--100 GeV; (7) Low limits on flux ratio between $G_{100}$ and $F^{upper}_{0.3-10}$.}
\end{table}

\end{document}